# IMPROVING HUMAN ANKLE JOINT POSITION SENSE

## USING AN ARTIFICIAL TONGUE-PLACED TACTILE BIOFEEDBACK

Nicolas VUILLERME, Olivier CHENU, Jacques DEMONGEOT and Yohan PAYAN

Laboratoire TIMC-IMAG, UMR CNRS 5525, La Tronche, France

Number of text pages of the whole manuscript (including figures and tables): 19

Number of figures : 3

Address for correspondence:

Nicolas VUILLERME

Laboratoire TIMC-IMAG, UMR CNRS 5525, Equipe AFIRM, Faculté de Médecine, 38706

La Tronche cédex, France.

Fax: (33) (0) 4 76 51 86 67

Email: nicolas.vuillerme@imag.fr

**Abstract**

Proprioception is comprised of sensory input from several sources including muscle spindles, joint capsule, ligaments and skin. The purpose of the present experiment was to investigate whether the central nervous system was able to integrate an artificial biofeedback delivered through electrotactile stimulation of the tongue to improve proprioceptive acuity at the ankle joint. To address this objective, nine young healthy adults were asked to perform an active ankle-matching task with and without biofeedback. The underlying principle of the biofeedback consisted of supplying subjects with supplementary information about the position of their matching ankle position relative to their reference ankle position through a tongue-placed tactile output device (Tongue Display Unit). Measures of the overall accuracy and the variability of the positioning were determined using the absolute error and the variable error, respectively. Results showed more accurate and more consistent matching performances with than without biofeedback, as indicated by decreased absolute and variables errors, respectively. These findings suggested that the central nervous system was able to take advantage of an artificial tongue-placed tactile biofeedback to improve the position sense at the ankle joint.

Accurate proprioception at the ankle joint is critical for body orientation and balance control and represents a prerequisite for different functional activities such as walking, running or driving. A recent study reporting a positive correlation between proprioceptive acuity ankle joint and postural control in stance [12], further suggested compromised ankle proprioception to be a predisposing factor for chronic ankle instability, balance difficulties, reduced mobility functions, fall, injury and re-injury [11,14,20,22,24].

It is generally agreed that proprioceptive sensation of a joint arises from various sensory receptors, termed *mechanoreceptors*, located in muscles, joints, tendons, ligaments and skin [13,19]. However, augmented/substituted sensory information also can be feed back to the central nervous system when one of the sensory proprioceptive inputs becomes unavailable/undermined/altered, or when one merely wants to enhance his proprioceptive acuity for accurate performances in daily-living, professional or sportive activities. The so-called *biofeedback* technique refers to the use of an external device to provide/increase an individual's awareness of sensory events that accompany performance. It is based on a closed-loop notion of motor control whereby errors in performances between intended action and outcome are detected and used to correct subsequent performances [21,28].

Augmented/substituted sensory biofeedback, widely used in physical therapy and rehabilitation, mostly is delivered though visual [18,38] or acoustic sensory channels (e.g., [10,15]). Interestingly, over the past decades, many efforts have been undertaken to develop human-machine interfaces (HMI) aimed at transmitting the information to the brain through tactile sense. Among them, the Tactile Vision Sensory Substitution (TVSS) system, developed by Bach-y-Rita and his collaborators, delivered visual information via arrays of stimulators in contact with the skin of one of several parts of the human body, including the abdomen, the chest, the back, the thigh, the brow, the fingertip [1-3,6,16,17]. Recently, these researchers developed a new HMI, the Tongue Display Unit (TDU), using the tongue as a substrate for electrotactile stimulation [2,4,5]. The choice to converge towards the electrostimulation of the tongue surface stems from three main reasons. On the one hand, the tongue being a very sensitive organ [4,30], largely represented within the somatosensory cortex [25] was shown to convey higher-resolution information than the skin can [27,31]. On the other hand, the tongue has an excellent conductance for electrical stimulation, hence requiring a lower voltage and current than those required for the skin to be stimulated [4]. Finally, the tongue is located in a protected environment of the mouth and is normally out of sight and out of the way.

While the TDU has proved its efficiency in the rehabilitative area in the context of visuo-tactile substitution designed to provide distal spatial information to blind individuals [2,5,6,27], it also can be considered as a relevant research tool to explore cognitive and brain mechanisms [2,5]. Along these lines, the purpose of the present experiment was to investigate whether the central nervous system was able to integrate an artificial biofeedback delivered through electrotactile stimulation of the tongue to improve proprioceptive acuity at the ankle joint. To address this objective, ankle joint position sense was evaluated using an active matching task, as recently done by others [11], with and without biofeedback.

Nine male young healthy males adults (age = $28.3 \pm 6.1$ years; body weight = $70.7 \pm 6.3$ kg; height = $179.7 \pm 5.6$ cm) voluntarily participated in the experiment. They gave their informed consent to the experimental procedure as required by the Helsinki declaration (1964) and the local Ethics Committee. None of the subjects presented any history of injury, surgery or pathology to either lower extremity that could affect their ability to perform the ankle joint position test.

Subjects were seated comfortably in a chair with their right and left foot secured to two rotating footplates. The knee joints were flexed at about $110°$. Movement was restricted to the ankle in the sagittal plane, with no movement occurring at the hip or knee. The axes of rotation of the footplates were aligned with the axes of rotation of the ankles. Precision linear potentiometers attached on both footplates provided analog voltage signals proportional to the ankles' angles. A handheld press-button allowed recording the matching. Signals from the potentiometers and the press-button were sampled at 100 Hz (12 bit A/D conversion), then processed and stored within the Labview 5.1 data acquisition system.

Subjects were barefoot for all testing, and care was taken to ensure that there were no discernible cues from the sole of the foot before testing. In addition, a panel was placed above the subject's legs to eliminate visual feedback about both ankles position.

The experimenter placed the left reference ankle at a predetermined angle where the position of the foot was maintained by means of a support [36]. Subjects therefore did not exert any effort to maintain the position of the left reference ankle, preventing the contribution of effort cues coming from the reference ankle to the sense of position during the test [36,37]. Two matching angular target positions were used: (1) 10° of plantarflexion ($P_{10°}$) and (2) 10° of dorsiflexion ($D_{10°}$). These positions were selected to avoid the extremes of the ankle range of motion to minimize additional sensory input from joint and cutaneous receptors [8]. Once the left foot had been positioned at the test angle ($P_{10°}$ vs. $D_{10°}$), subject's task was to match its position by voluntary placement of their right leg. When they felt that they had reached the target angular position (i.e., when the right foot was presumably aligned with the left foot), they were asked to press the button held in their right hand, thereby registering the matched position. This active matching task was performed under two No-biofeedback and Biofeedback experimental conditions. The No-biofeedback condition served as a control condition. In the Biofeedback condition, subjects performed the task using a TDU-biofeedback system. The underlying principle consisted of supplying subjects with supplementary biofeedback about the position of the matching right ankle relative to the reference left ankle position through a tongue-placed tactile output device. Electrotactile stimuli were delivered to the front part of the tongue dorsum via flexible electrode arrays placed in the mouth, with connection to the stimulator apparatus via a flat cable passing out of the mouth. The system used in the present experiment comprises a 2D electrode array (1.5 × 1.5 cm) consisting of 36 gold-plated contacts each with a 1.4 mm diameter, arranged in a 6 × 6 matrix (Figure 1).

-------------------------------------

Please insert Figure 1 about here

-------------------------------------

The following coding scheme for the TDU was used: (1) no electrical stimulation when both ankles were in a similar angular position within a range of 1.5° (Figure 2A); (2) stimulation of either the anterior or posterior zone of the matrix (2 × 6 matrix) (i.e. stimulation of front and rear portions of the tongue) depending on whether the matching right ankle was in a too plantarflexed or dorsiflexed position relative to the reference left ankle, respectively (Figures 2B and 2C, respectively). The frequency of the stimulation was maintained constant at 50 Hz across subjects, ensuring a sensation of the continuous stimulation over the tongue surface. Conversely, the sensitivity to the electrotactile stimulation varying not only between individuals but also as a function of location on the tongue, as reported in a preliminary experiment, the intensity of the electrical stimulating current was adjusted for each subject, and each of the front and rear portions of the tongue.

-------------------------------------

Please insert Figure 2 about here

-------------------------------------

Several practice runs were performed prior to the test to ensure that subjects had mastered the relationship between ankle angular positions and lingual stimulations and to gain confidence with the TDU.

Five trials for each target angular position and each experimental condition were performed, representing a total of 20 trials per subject. The order of presentation of the two targets angular positions ($P_{10°}$ *vs.* $D_{10°}$) and the two experimental conditions (No-biofeedback *vs.* Biofeedback) was randomized. Subjects were not given feedback about their performance and errors in the position of the right ankle were not corrected.

Two dependent variables were used to assess matching performances [28]. (1) The absolute error (AE), the absolute value of the difference between the position of the right matching ankle and the position of the left reference ankle, is a measure of the overall accuracy of positioning. (2) The variable error (VE), the variance around the mean constant error score, is a measure of the variability of the positioning. Decreased AE and VE scores indicate increased accuracy and consistency of the positioning, respectively [28].

The means of the five trials performed in each of the four experimental conditions were used for statistical analyses. 2 Conditions (No-biofeedback *vs.* Biofeedback) × 2 Targets angular positions ($P_{10°}$ *vs.* $D_{10°}$) analyses of variances (ANOVAs) with repeated measures of both factors were applied to the AE and VE data. Level of significance was set at 0.05.

Analysis of the AE showed a main effect of Condition, yielding smaller values in the Biofeedback than No-biofeedback condition ($F(1,8) = 29.97$, $P < 0.001$, Figure 3A). The ANOVAs showed no main effect of Target angular position, nor any interaction of Condition × Target angular position ($P > 0.05$).

Analysis of the VE also showed a main effect of Condition yielding smaller values in the Biofeedback than No-biofeedback condition ($F(1,8) = 16.85$, $P < 0.01$, Figure 3B). The ANOVAs showed no main effect of Target angular position, nor any interaction of Condition × Target angular position ($Ps > 0.05$).

------------------------------------

Please insert Figure 3 about here

------------------------------------

The purpose of the present experiment was to investigate whether the central nervous system was able to integrate an artificial biofeedback delivered through electrotactile stimulation of the tongue to improve proprioceptive acuity at the ankle joint. To address this objective, nine young healthy adults were asked to perform an active ankle-matching task. This task was executed for two angular target positions of (1) 10° of plantarflexion ($P_{10°}$) and (2) 10° of dorsiflexion ($D_{10°}$) and for two experimental conditions of (1) No-biofeedback and (2) Biofeedback. In the latter condition, position of the matching ankle relative to the reference ankle was fed back to a tongue-placed tactile output device (TDU) generating electrotactile stimulation on a 36-point ($6 \times 6$) matrix held against the surface of the tongue dorsum (Figure 1). It is important to emphasize that all subjects reported that the biofeedback was comfortable and its way of representing the information was intuitive. Measures of the overall accuracy and the variability of the positioning were determined using the AE and the VE, respectively [28].

Reduced AE (Figure 3A) and VE (Figure 3B) scores were observed in the Biofeedback relative to the No-biofeedback condition. In addition to the absence of significant interaction of Condition $\times$ Target angular position, these results showed more accurate and more consistent matching performances, whatever the target angular position, when biofeedback was in use than when it was not. In other words, from a neuroscience perspective, these finding suggested that the central nervous system was able to take advantage of an artificial tongue-placed tactile biofeedback to improve the position sense at the ankle joint.

At this point, it is conceivable that the multisensory integration, referring the central nervous system's capacity of combining information coming from different sensory modalities to provide a more accurate representation of the body [9], requires cognitive resources. This integration process is likely to be all the more cognitive demanding because an unusual source of information, such as that used in our Biofeedback condition, has to be integrated by the central nervous system. Although the experimental task and the augmented sensory information were different, this assumption is reminiscent of a recent report suggesting that the ability to use a light fingertip touch as a source of sensory information to improve postural control during bipedal quiet standing is attention demanding [35]. Along these lines, it is possible that the availability of the artificial tongue-placed tactile biofeedback, allowing young healthy adults to improve their ankle joint position sense (Figure 3), could lead different effects when engaging in concurrent cognitive tasks or with individuals suffering from cognitive deficits (e.g., elderly persons, stroke patients). Such a proposal is yet speculative and warrants additional investigations. Finally, the present results evidenced the effectiveness of a TDU-biofeedback system for improving ankle joint position sense under "normal" proprioceptive conditions (i.e., with redundant and reliable sensory information). Whether the central nervous system would able to integrate this source of information to compensate for a degradation of proprioceptive input consecutive to either experimental manipulations [11,23,26], trauma [7,14,24], normal aging [20,34], or disease [29,32,33] was not addressed in the present study and also merits further research. We believe that such a perspective not only would be of great theorical value in terms of sensory reweigthing mechanisms involved in motor control, by documenting how the central nervous system integrates an artificial biofeedback and combines it with natural sensory cues, but also could have relevant implications in rehabilitative and ergonomical areas by assisting disabled individuals for whom the consequences of ankle positioning error could be more dramatic (e.g., elderly persons, patients with diabetic neuropathy).

**Acknowledgements**

The authors are indebted to Professor Paul Bach-y-Rita for introducing us to the TDU and for discussions about sensory substitution. The authors would like to thank subject volunteers. Special thanks also are extended to Damien Flammarion and Sylvain Maubleu for technical assistance and Louis V. for various contributions.

**Figure captions**

**Figure 1.** Photograph of the Tongue Display Unit used in the present experiment. It comprises a 2D electrode array ($1.5 \times 1.5$ cm) consisting of 36 gold-plated contacts each with a 1.4 mm diameter, arranged in a $6 \times 6$ matrix.

**Figure 2.** Sensory coding schemes for the TDU (*lower panels*) as a function of the position of the matching right ankle relative to the reference left ankle (*upper panels*). Black dots represent activated electrodes. There were three possible stimulation patterns of the TDU:

**A**: no electrodes activated when both ankles were in a similar angular position within a range of 1.5°.

**B**: 12 electrodes ($2 \times 6$) of the anterior zone of the matrix are activated (corresponding to the stimulation of the front portion of the tongue dorsum) when the matching right ankle was in a too plantarflexed position relative to the reference left ankle.

**C**: 12 electrodes ($2 \times 6$) of the posterior zone of the matrix are activated (corresponding to the stimulation of the rear portion of the tongue dorsum) when the matching right ankle was in a too dorsiflexed position relative to the reference left ankle.

**Figure 3.** Mean and standard deviation for the absolute error (A) and the variable error (B) for the two No-biofeedback and Biofeedback conditions. The two experimental conditions are presented with different symbols: No-biofeedback (*white bars*) and Biofeedback (*black bars*). The significant *P*-values for comparison between No-biofeedback and Biofeedback conditions also are reported (**: $P<0.01$, ***: $P<0.001$).

**Figure 1**

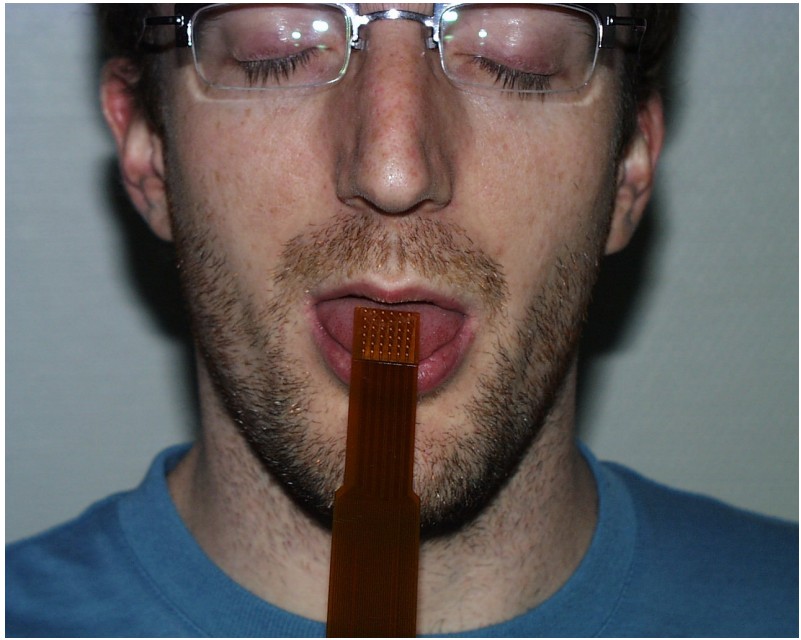

**Figure 2**

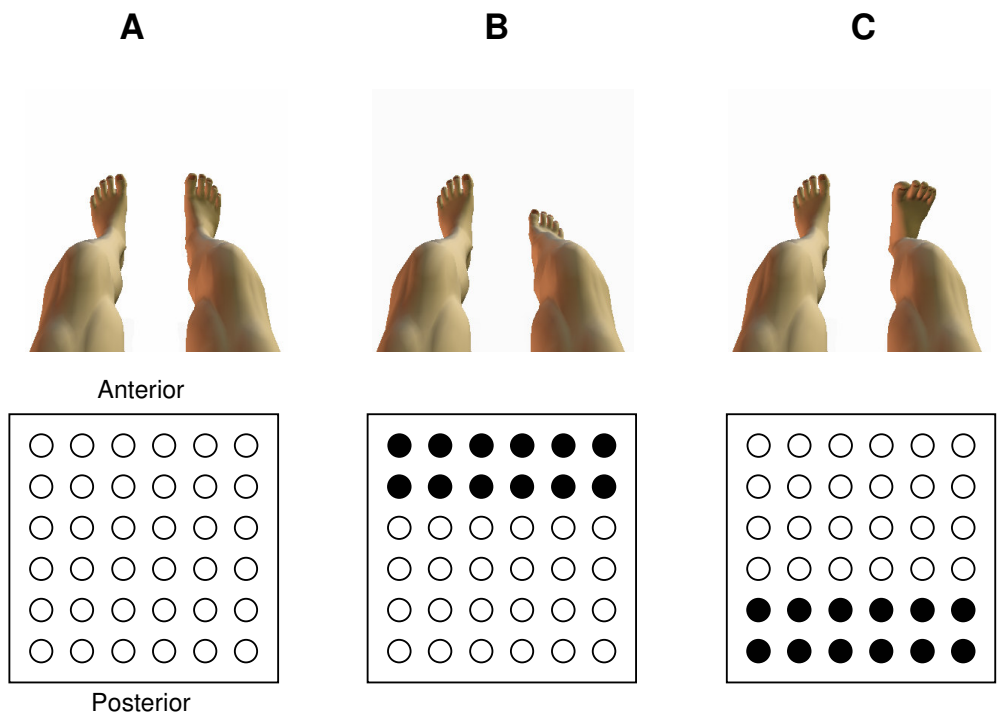

A             B             C

**Figure 3**

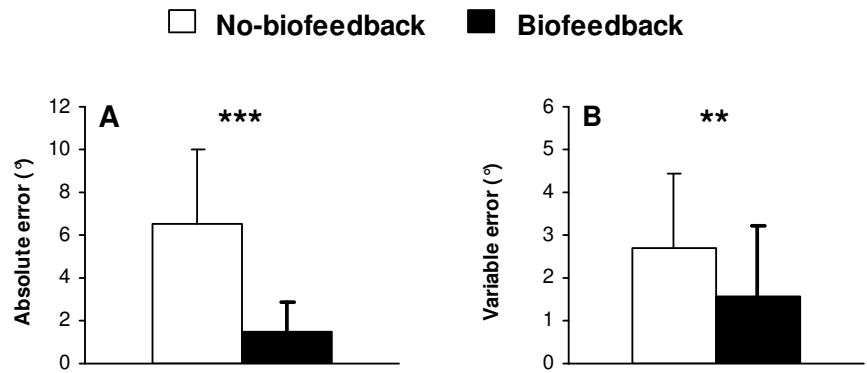